**The Importance of Conference Proceedings in Research Evaluation: a Methodology for Assessing Conference Impact**

Dmitry Kochetkov[1,2*], Aliaksandr Birukou[3, 4], Anna Ermolayeva[3]

[1] Centre for Science and Technology Studies, Leiden University, Leiden, The Netherlands

[2] Higher School of Economics, Moscow, Russia

[3] Peoples' Friendship University of Russia (RUDN University), Moscow, Russia

[4] Springer Nature, Heidelberg, Germany

* Corresponding author. E-mail: d.kochetkov@cwts.leidenuniv.nl

**Abstract.** Conferences are an essential tool for scientific communication. In disciplines such as Computer Science, over 50% of original research results are published in conference proceedings. In this study, we have analyzed the role of conference proceedings in various disciplines and propose an alternative approach to research evaluation based on conference proceedings and Scimago Journal Rank (SJR). The result of the study is a list of conference proceedings, categorized Q1 - Q4 in several disciplines by analogy with SJR journal quartiles. The comparison of this bibliometric-driven ranking with the expert-driven CORE ranking in Computer Science showed a 62% overlap, as well as a significant average rank correlation of the category distribution. Moreover, 38 conference proceedings in Engineering (45% of the list) and 23 in Computer Science (32% of the list) have an SJR level corresponding to the first quartile journals in these areas. This again emphasizes the exceptional importance of conferences in these disciplines.

**Introduction**

In China, India (Madhan et. al, 2018), Russia, Turkey[1], UK (Koya and Chowdhury, 2017) and other countries, research evaluation is based on the indicators of the sources, i.e., journals, conference proceedings, book series, in which the results are published. This often leads to labeling publication sources with several predefined classes and judging the importance of a publication based on the class of the source. As in many research areas original results are published in journals (Vrettas and Sanderson 2015), research evaluation policies are often biased towards journals.

---

[1] https://www.urapcenter.org/Methodology, last accessed 03.07.2020

As an example of such policy, we can mention the ongoing discussion on the Comprehensive Methodology for Evaluating Publication Performance (CMEPP) in Russia. The methodology is based on the evaluation of publications depending on the Impact Factor (IF) quartile of journals in Web of Science: a journal article published in the first quartile journal gets 20 points, the second quartile - 10 points, the third quartile - 5 points, and the fourth quartile - 2.5 points. All other publications, including conference papers, book chapters, journal articles, indexed only in Scopus or Russian Science Citation Index, receive 1 point. This scale applies to the natural sciences, engineering, and life sciences. There is a flat scale for social sciences and humanities: all publications in Scopus or Web of Science get 3 points regardless of the quartile. The advantages of this technique include its simplicity and the ability to evaluate the research manuscript directly at the time of publication, as the process of gathering citations takes time). The disadvantages of the method are discussed in the Declaration on Research Assessment, DORA (Cagan, 2013) and include the possibility of manipulation with quantitative journal metrics (The PLoS Medicine Editors, 2006), although this is true of all purely quantitative methods of evaluation. However, the main drawbacks of the methodology are the possible discrepancy between the citation count of a specific article and the value of the journal where it is published and high variance in the citations to the articles in the same journal (Seglen, 1997). More recent research (Waltman & Traag, 2020) shows that it is possible that the Impact Factor is more accurate predictor of the value of an article than the number of citations it has received. However, the design of the study does not allow the authors to make conclusions about whether in practice the Impact Factor is indeed more accurate than the number of citations. Without going into further details of article-level vs journal-level metrics arguments, we will try to reconcile the debates for the purpose of this paper. IF and other journal-level metrics are widely used in research evaluation to judge the value of research contributions or researchers. If an article was published in a Q1 journal, it most probably has a higher probability of being superb than an article in a Q4 journal in the same discipline. However, counter-examples could be found, i.e., a weak article in the Q1 journal or a superb article in the Q4 journal. Therefore, article-level metrics, or even better, qualitative research methods should be used next to the quantitative journal-level indicators. Now, let us turn back to the purpose of the paper and stress that the CMEPP approach, as well as most of journal-level approaches neglects the conferences: even the most prestigious conferences receive a lower rating than articles in the journals of the fourth quartile.

In this paper we 1) review the current practices of using conferences in the research evaluation; 2) identify scientific disciplines, where conference proceedings play a significant role in the communication of primary research results; 3) propose a new methodology for the assessment of conference proceedings based on Scopus and Scimago Journal Rank (SJR) data; 4) show that such bibliometric-driven methodology produces classification of conferences similar to the classification designed by domain experts, such as CORE.

**The role of conferences**

Conferences play an incredibly significant role in some areas of science - for example, in *Computer Science*, more than 60% of the research results are published in conference proceedings (Meho, 2019). Vrettas & Sanderson, 2015 show that a small number of elite conferences have higher average citations rate than elite journals. While there is a range of metrics for evaluating journals, e.g, *Impact Factor*, SCImago Journal Rank (*SJR)*, there is no common metric for evaluating conferences (Almendra, Enăchescu, & Enăchescu, 2015).

The first attempts to create conference rankings appeared in Computer Science. The most frequently used rankings, based on the experience of the authors, are the *Computing Research and Education Association of Australasia*, *CORE* and *ERA Ranking* (Butler, 2008), *Qualis* (2012), China Computer Federation (*CCF*) ranking (2012), *MSAR* (2014), and *GII-GRIN-SCIE* (2014). The *CORE Conference Ranking*[2] provides evaluation of major computer science conferences. Decisions are made by academic committees based on data requested as part of the submission process. ERA Ranking (2010) was created in the framework of *Excellence in Research in Australia (ERA)*[3]. It incorporates data from a previous ranking attempt done by *CORE*. *Qualis*[4] has been published by the Brazilian ministry of education and uses the H-index (strictly speaking, H-index percentiles) as performance measure for conferences. The latest edition was released in 2016. The fifth edition of the list of journals and conferences recommended by the CCF[5] was released in April 2019 and assigns A, B, C ranks to conferences in each CS subarea. Interestingly, the release mentions that "the influence of journals and conferences is not directly related to the influence of a single paper published there. Therefore, it is not recommended to use this list as a basis for academic evaluation" – see our earlier discussion on journal- vs. article-level metrics. *MSAR*[6] is the *Microsoft Academic's* field rankings for conferences. It is similar to the h-index and calculates the number of publications by an author and the distribution of citations to the publications.

---

[2] Available at http://www.core.edu.au/conference-portal (date accessed 01.03.2020).
[3] Available at https://www.arc.gov.au/excellence-research-australia (date accessed 01.03.2020).
[4] The official version (latest ranking of conferences "Conferências" for 2013-2016) is available at https://sucupira.capes.gov.br/sucupira/. However, the tool at UFMT (Universidade Federal de Mato Grosso, available at https://qualis.ic.ufmt.br/, or the complete list available at https://www.capes.gov.br/images/documentos/Qualis_periodicos_2016/Qualis_conferencia_ccomp.pdf are easier to use (date accessed 28.04.2020).
[5] Available at https://www.ccf.org.cn/c/2019-04-25/663625.shtml (date accessed 28.04.2020).
[6] Available at https://academic.microsoft.com/home (date accessed 04.03.2020).

Field ranking only calculates publications and citations within a specific field and shows the impact of the scholar or journal within that specific field. GII-GRIN-SCIE conference rating[7] is an attempt to develop a unified rating of computer science conferences led by GII (Group of Italian Professors of Computer Engineering), GRIN (Group of Italian Professors of Computer Science), and SCIE (Spanish Computer-Science Society), see Cabitza and Locoro, 2015. The latest version ranks all conferences in four tiers (top notch, very high quality, good, and work in progress) based on the CORE, Qualis and MSAR rankings and it was released in 2018. Last but not the least, there is also Google Scholar Top publications[8] tool which lists both conference and journals in the same ranking.

Now that we reviewed why conferences are important in Computer Science and how this is tackled with different rankings, let us check if conferences only matter in Computer Science. We used Scopus to compute the share of conference proceedings in the total number of publications for particular subject categories in 2015-2019. Figure 1 shows the subject categories where the share of conference proceedings is more than 10% of all publication sources. The highest percentage of conferences is observed in *Computer Science*, as well as *Mathematics* and *Decision Sciences*. Note that given that publications in Scopus can belong to several categories, there might be overlap. There is a substantial share of publications in conference proceedings for *Engineering* and *Energy*.

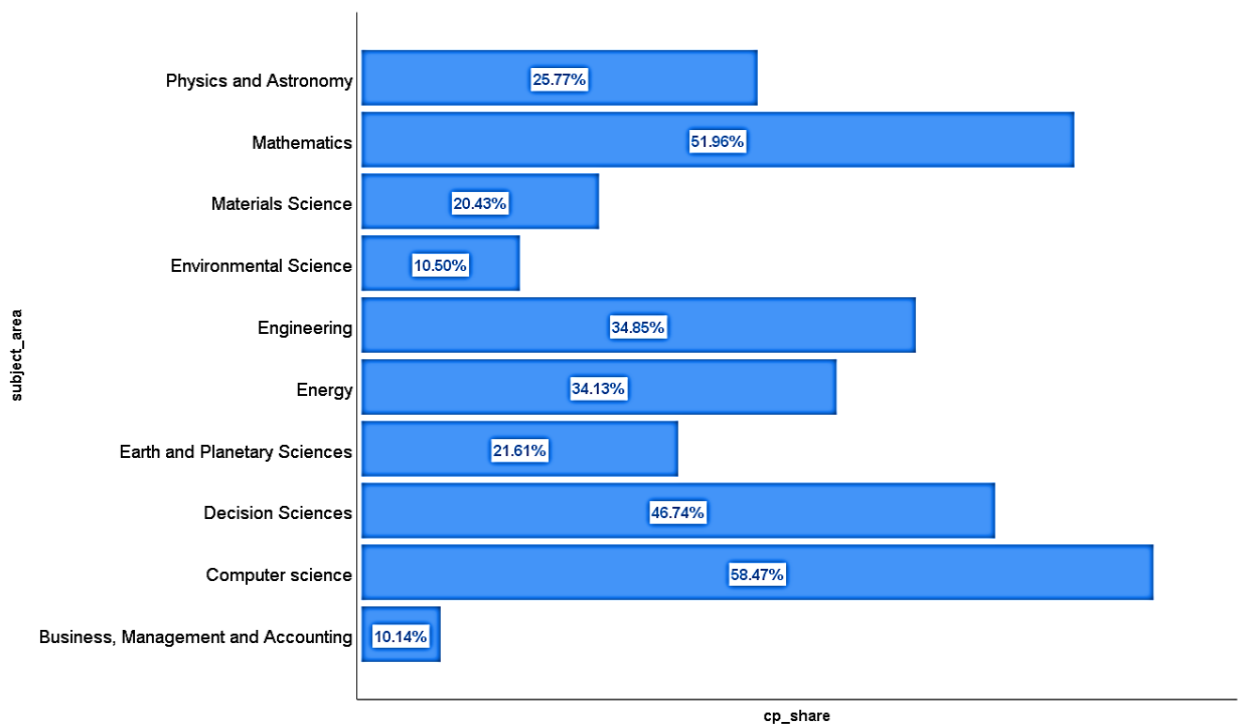

---

[7] Available at http://gii-grin-scie-rating.scie.es/ (date accessed 28.04.2020)
[8] https://scholar.google.com/citations?view_op=top_venues&hl=en (date accessed 21.06.2020)

Fig. 1 The share of conference proceedings in the total number of publications for particular subject categories 2015-2019. Source: authors' own calculations based on *Scopus*[9] data (data retrieved on 19.03.2020)

The conference rankings mentioned above provide an initial basis for the evaluation of research published in conference proceedings. However, a lot of them have been created for specific local communities and were never meant to be used globally, like journal IFs. While most of the current rankings are based on citations, alternative metrics can be used to identify influential articles and authors, e.g., *PageRank*-like algorithms (Dunaiski, Visser, & Geldenhuys, 2016). Taking into account the phenomenon of predatory or fake conferences (Bilder, 2017), one could create a white list of conferences and criteria to assess the contribution of specific conferences to academic results and reputation. In the following, we will propose a metric for the conferences indexed in Scopus and compare it with the most frequently used CS conference ranking, CORE.

**Data and Methods**

At the first stage, we selected conference proceedings included in the *Scopus* sources list[10] (*Conference Proceedings post-1995*), which are currently indexed (i.e., have the *ongoing* status), and for which an *SJR*[11] score is available. This selection resulted in 171 conference proceedings. The *SCImago Journal Rank (SJR)* is not just a citation indicator such as *Impact Factor* or *CiteScore*; it is based on a *PageRank*-like algorithm, which is an iterative process of prestige transfer among the publication sources. The calculation is an iterative process in which the prestige of each source depends on the prestige of the sources which cite it. The final *SJR* value is normalized over the number of documents published in the citation window (SCImago, 2007). Given that the *SJR* is computed based on *Scopus* data, we also used this database in our analysis. Out of the 171 conference proceedings, 153 were assigned one or more subject categories (*third level ASJC*[12]) in Scopus. For the 18 conferences that were not assigned any subject category; we deduced the categories based on publications in *Scopus*.

The distribution of conferences across subject categories is shown in Fig. 2. Note that one conference can belong to several subject categories. Out of 171 conferences, one was assigned to five subject categories, one to four, 13 to three, 66 to two, and 90 conferences had only one subject category.

---

[9] Available at https://www.scopus.com/ (date accessed 19.03.2020).
[10] Available at https://www.scopus.com/ (date accessed 04.03.2020).
[11] Available at https://www.scimagojr.com/ (date accessed 04.03.2020).
[12] All Science Journal Classification Codes. Available at https://service.elsevier.com/app/answers/detail/a_id/15181/supporthub/scopus/ (date accessed 04.03.2020).

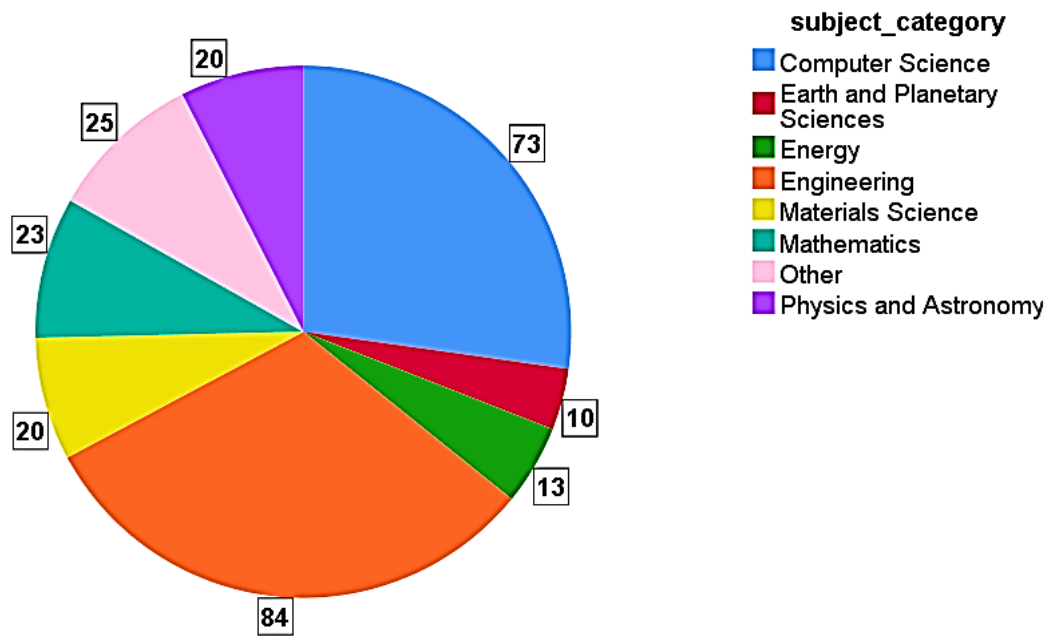

Fig. 2 The distribution of conferences across subject categories. Source: authors' own calculations

Next, for each of the subject category, we computed the threshold *SJR* values for the quartiles, in the same way SCImago calculates them for journals. This was necessary because SCImago does not assign quartiles to the conferences, only to journals and book series. This allowed us to assign each conference to the corresponding quartile (*Q1, Q2, Q3, Q4*) in each subject category. For example, the minimum SJR for journals and book series of the first quartile is 0.261, the second is 0.139, the third is 0.104, and the fourth is 0.1. The IOP Conference Series: Materials Science and Engineering has an SJR of 0.195, so we can classify the source as Q2. We emphasize that this is not a quartile itself; it is a conditional assignment of conference proceedings to a quartile based on the SJR value. For journals covering several subjects, CMEPP research evaluation guidelines suggest using the maximum of the quartiles in those subjects. However, the importance of the same conference in different communities varies, as also mentioned in the CCF release notes. We therefore would like to stress the importance of using subject-specific quartiles for conferences, i.e., a conference can belong to several subject categories and can have different quartiles there.

**Results and Discussion**

Figure 3 shows the the distribution of conferences across quartiles in the context of subject categories. If one considers all sources (journals, book series, conference proceedings), the share of each quartile is obviously 25%. However, as our selection is limited to the conference proceedings, the distribution between the categories Q1-Q4, is very different for each subject category.

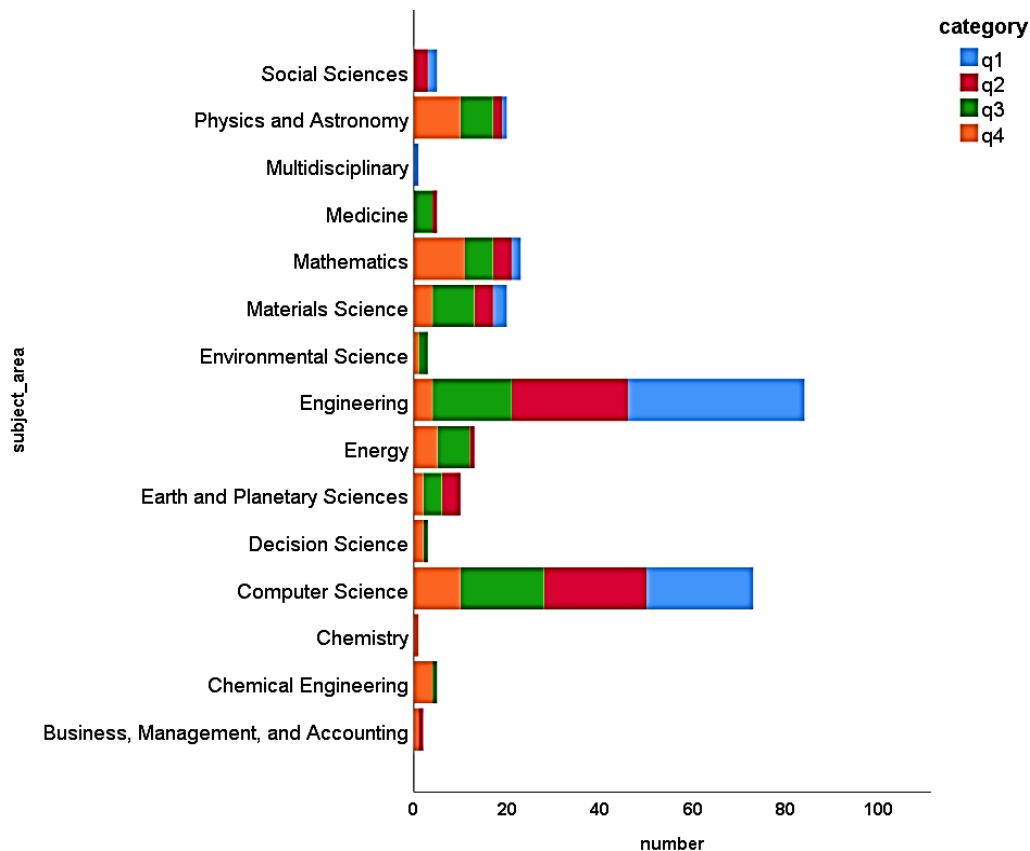

**Fig. 3** Distribution of conference proceedings into categories. Source: authors' own calculations

From the graph, it is evident that Engineering and Computer Science have not only the highest share of conferences but also the largest number of high-impact conferences. This once again confirms the thesis that conference proceedings must be considered when evaluating research in these areas. Our results are in line with with the results of earlier studies (Meho, 2019). The difference in the number and quality of conferences between these subject categories and the rest is substantial. The source data is available in (Kochetkov et al., 2020).

Out of the 73 proceedings of Computer Science conferences, 45 conferences (62%) are in the CORE ranking; 10 sources are aggregators that publish the proceedings of many conferences (e.g., Procedia Computer Science, ACM International Conference Proceeding Series, etc.); and 18 conferences are not core CS conferences, but are from related fields (for example, IEEE MTT-S International Microwave Symposium Digest). The latter appear in our dataset because according to the ASJC classification conferences can fall simultaneously into several subject areas/categories. Such conferences, however, are out of scope for CORE, which focuses exclusively on Computer Science conferences.

For the 45 conferences from our list, which are also present in CORE, we compared the distribution by category (Fig. 4, Q1 for SJR corresponds to A * for CORE, Q2 - A, Q3 - B, Q4 - C). Spearman's rank correlation coefficient was 0.452, which suggests an average correlation dependence. This is an interesting fact, given the fundamentally different approaches to the formation of lists, bibliometric and expert. The full table is also presented in the dataset available online (see footnote 12).

**Table 1:** A Comparative Analysis of Distribution of Conferences into Categories*

| SJR 2018/CORE 2018 | A* | A | B | C | NA |
|---|---|---|---|---|---|
| Q1 | 11 | 4 | 4 | 1 | 3 |
| Q2 | 5 | 7 | 2 | 1 | 7 |
| Q3 | - | 2 | 6 | 1 | 9 |
| Q4 | - | - | - | 1 | 9 |
| NA | 51 | 407 | 402 | 793 | |

*Source: author's calculations.

The study has several limitations:

1. We have evaluated conference proceedings, not the conferences themselves. If one would like to evaluate conferences, they should take into account not only the bibliometric data, but also various other parameters: topical scope, program committee, authors, the peer review process, proceedings publication culture, etc. However, a quantitative assessment presented here may be a convenient auxiliary tool, even though it does not eliminate the need for expert evaluation.

2. The list reflects only non-journal and non-book sources. Conference proceedings published in journals and book series (e.g., Journal of Physics Conference Series, Lecture Notes in Computer Science) can use the SJR quartile of the corresponding journal or the book series.

3. The list only reflects conference proceedings with the serial ISSN; some conferences do not receive it due to the oversight of the organizing committee. Such conferences are not included in the list of serials in Scopus and could not be included in the analysis.

4. The list includes not only the proceedings of individual conferences but also aggregators such as CEUR Workshop Proceedings, Leibniz International Proceedings in Informatics (LIPIcs). The level of conferences within such publications may vary significantly. Unfortunately, the data granularity in Scopus does not allow for the conference-level analysis within these sources.

Even though the CCF recommends not using conference rankings for academic evaluation, Li et al. (2018) show how such rankings influence publishing behavior of scientists. Therefore, it is important to provide more transparency in how rankings are created, what is included, which metrics are used, etc. The methodology proposed in this paper represents a step in this direction, as it combines transparent bibliometric indicators and correlates with expert opinions.

The authors will continue research on the evaluation of conferences and conference papers. We would like to move towards paper-level metrics, as different papers in the same conference proceedings have different quality, citations, importance. In this regard we would like to mention the ConfRef.org project, which was created to provide information on scientific conferences and provide standard identifiers for conferences. The current prototype provides data on 40,000 conferences, mainly from computer science, provided by Springer Nature and DBLP. The primary purpose of ConfRef is to provide trusted information about the history, dates, venues, places of publication / past issues of a series of conferences (and related conferences) in various disciplines (Computer Science, Electrical Engineering, Mathematics), as well as information about upcoming conferences and invitations, dates and information about program committee. On top of this, ConfRef will deal with identifying predatory or fake conferences.

**Conclusion**

In this paper we made an attempt to review the role of conferences in the research evaluation and to identify scientific disciplines, where conference proceedings are an important outlet for publishing original research results. Next to the "usual suspects", i.e. Computer Science, conference proceedings often used for publishing results in Engineering, Mathematics, Energy, Decision Sciences. We also presented a new methodology for applying Scopus and Scimago Journal Rank (SJR) data for the assessment of conference proceedings and showed that it provides similar results to expert-designed ranking, such as CORE. The methodology shows that some conference proceedings in Computer Science, Engineering, Material Science, Physics and Astronomy, and Mathematics are comparable with Q1-Q2 journals.

Future work includes development of tools which would implement the proposed methodology and work on removing the limitations such as the different granularity of conference proceedings sources.

**Acknowledgements**

We are grateful to Volha Bryl, Andrey Gromyko, and Kai Eckert from the ConfRef.org project, to Ludo Waltman, Vincent Traag and all the members of QSS group of CWTS, Leiden University, and to the participants of the Crossref/DataCite working group on persistent identifiers for conferences[13] for useful discussions.

---

[13] https://www.crossref.org/working-groups/conferences-projects/

**Funding**

The publication has been prepared with the support of the "RUDN University Program 5-100" (recipient A. Birukou). The study was also supported by the Russian Foundation for Basic Research, project 18-00-01040 KOMFI "The Impact of Emerging Technologies on Urban Environment and the Quality of Life of Urban Communities" (recipient D. Kochetkov).

**Conflict of Interest**

Aliaksandr Birukou is employed by Springer Nature and was previously responsible for computer science proceedings program.

**References**

Almendra, V. da S., Enăchescu, D., & Enăchescu, C. (2015). Ranking computer science conferences using self-organizing maps with dynamic node splitting. *Scientometrics*, *102*(1), 267–283. https://doi.org/10.1007/s11192-014-1436-y

Bilder, G. (2017). Taking the "con" out of conferences. Crossref blog. https://www.crossref.org/blog/taking-the-con-out-of-conferences/

Butler, L. (2008). ICT assessment: Moving beyond journal outputs. *Scientometrics*, *74*(1), 39–55. https://doi.org/10.1007/s11192-008-0102-7

Cabitza, F., Locoro, A. (2015). Exploiting the Collective Knowledge of Communities of Experts - The Case of Conference Ranking. *Proceedings of the 7th International Joint Conference on Knowledge Discovery, Knowledge Engineering and Knowledge Management (KMIS 2015),* vol. 3, 159-167. https://doi.org/10.5220/0005592601590167

Cagan, R. (2013). The San Francisco Declaration on Research Assessment. *DMM Disease Models and Mechanisms*, 869-870, 6(4). https://doi.org/10.1242/dmm.012955

Dunaiski, M., Visser, W., & Geldenhuys, J. (2016). Evaluating paper and author ranking algorithms using impact and contribution awards. *Journal of Informetrics*, *10*(2), 392–407. https://doi.org/10.1016/j.joi.2016.01.010

Kochetkov, Dmitry; Birukou, Aliaksandr ; Ermolayeva, Anna (2020), "Methodology for Conference Proceedings Assessment: a Conference Proceedings Dataset", Mendeley Data, v5 http://dx.doi.org/10.17632/hswn9y67rn.5

Koya K, Chowdhury G (2017) Metric-based vs peer-reviewed evaluation of a research output: Lesson learnt from UK's national research assessment exercise. PLoS ONE 12(7): e0179722. https://doi.org/10.1371/journal.pone.0179722

Li, X., Rong, W., Shi, H., Tang, J., Xiong, Z. (2018) The impact of conference ranking systems in computer science: a comparative regression analysis. *Scientometrics* **116,** 879–907. https://doi.org/10.1007/s11192-018-2763-1

Madhan, Muthu ; Gunasekaran, Subbiah ; Arunachalam, Subbiah Evaluation of research in India – are we doing it right?. *Indian Journal of Medical Ethics*, [S.l.], v. 3, n. 3 (NS), p. 221, mar. 2018. https://doi.org/10.20529/IJME.2018.024

Meho, L. I. (2019). Using Scopus's CiteScore for assessing the quality of computer science conferences. *Journal of Informetrics*, *13*(1), 419–433. https://doi.org/10.1016/j.joi.2019.02.006

SCImago. (2007). Description of Scimago Journal Rank Indicator. Retrieved from https://www.scimagojr.com/SCImagoJournalRank.pdf

Seglen, P.O. (1997) Why the impact factor of journals should not be used for evaluating research. BMJ 314, 498–502. https://doi.org/10.1136/bmj.314.7079.497

The PLoS Medicine Editors (2006). The impact factor game. PLoS Med 3(6): e291 https://doi.org/10.1371/journal.pmed.0030291

Vrettas, G., & Sanderson, M. (2015). Conferences versus journals in computer science. *Journal of the Association for Information Science and Technology*, *66*(12), 2674–2684. https://doi.org/10.1002/asi.23349

Waltman, L., & Traag, V. A. (2020). Use of the journal impact factor for assessing individual articles need not be statistically wrong. *F1000Research*, *9*, 366. https://doi.org/10.12688/f1000research.23418.1